\documentclass{acm_proc_article-sp}

\usepackage{hyperref}

\renewcommand\footnoterule{\rule{\linewidth}{0.2pt}}
\renewcommand{\footnotesize}{\scriptsize}

\begin{document}

\title{Lockdown: Dynamic Control-Flow Integrity}

\numberofauthors{3} %
\author{
\alignauthor
Mathias Payer\\
       \affaddr{Purdue University, USA}\\
\alignauthor
Antonio Barresi\\
       \affaddr{ETH Zurich, Switzerland}\\
\alignauthor
Thomas R. Gross\\
       \affaddr{ETH Zurich, Switzerland}\\
}

\maketitle
\begin{abstract}
Applications written in low-level languages without type or memory safety
are especially prone to memory corruption. Attackers gain code execution
capabilities through such applications despite all currently deployed defenses
by exploiting memory corruption vulnerabilities. Control-Flow Integrity (CFI) is
a promising defense mechanism that restricts open control-flow transfers to a
static set of well-known locations.\\
We present Lockdown, an approach to dynamic CFI that protects legacy,
binary-only executables and libraries. %
Lockdown adaptively learns the control-flow
graph of a running process using information from a trusted dynamic loader.  The
sandbox component of Lockdown restricts interactions between different
shared objects to imported and exported functions by enforcing fine-grained
CFI checks. Our prototype implementation shows that dynamic CFI results in low
performance overhead. %

\end{abstract}

\section{Introduction}

\let\thefootnote\relax\footnotetext{
\\
Technical Report, Laboratory for Software Technology, ETH Zurich\\
Copyright 2014 ETH Zurich\\
http://dx.doi.org/10.3929/ethz-a-010171214}

Memory corruption is a well-known problem for applications written in low-level
languages that do not support memory safety or type safety (e.g., C, or C++). The
core of many applications running on current systems is written in C or C++, and
it is simply impossible to rewrite all these applications in a safe language due
to the large amount of existing code. In addition, the problem of memory
corruption is not restricted to low-level languages as safe languages are often
implemented using low-level languages (e.g., the HotSpot Java virtual machine is
implemented in C++) or use low-level runtime libraries like the libc. Since
2006, a number of defense mechanisms like Address Space Layout Randomization
(ASLR)~\cite{aslr}, Data Execution Prevention (DEP)~\cite{execshield}, stack
canaries~\cite{propolice01hiroaki}, and safe exception handlers have been
deployed in practice to limit the power of attacker-controlled memory
corruption. Unfortunately, all commonly deployed defense mechanisms can be
circumvented. The list of Common Vulnerabilities and Exposures (CVE) shows that
(i) memory corruption is common and (ii) many of these memory corruption
vulnerabilities can be used to redirect the control-flow of the application to
execute attacker-controlled code (either by injecting new code or by reusing
existing code sequences in an unintended way). 
Control-Flow Integrity (CFI)~\cite{ccs05erlingsson, osdi06erlingsson,
philippaerts11dimva, bletsch11acsac, wang10oakland, zeng13usenix,
zhang13oakland, zhang13asiaccs, zhang13security, criswell14sp, niu14pldi} is a
promising defense mechanism that, by design, restricts the set of targets that
can be reached by any control-flow transfer according to the statically
determined control-flow graph. Unfortunately, current implementations share one
or more of the following drawbacks: (i) some imprecision in the protection by
allowing a larger set of targets than originally possible with different levels
of imprecision (this imprecision can be exploited by new
attacks~\cite{bittau14sp, bosman14sp, goektas14sp}), (ii) the need to recompile
applications~\cite{ccs05erlingsson, osdi06erlingsson, philippaerts11dimva,
bletsch11acsac, wang10oakland, zeng13usenix}, (iii) no support (or protection)
for shared libraries~\cite{ccs05erlingsson, osdi06erlingsson,
philippaerts11dimva, bletsch11acsac, wang10oakland}, or (iv) relying on precise
relocation information that is only available in Windows
binaries~\cite{zhang13oakland, zhang13asiaccs}.

\begin{figure}[t]
  \begin{center}
    \includegraphics[width=85mm]{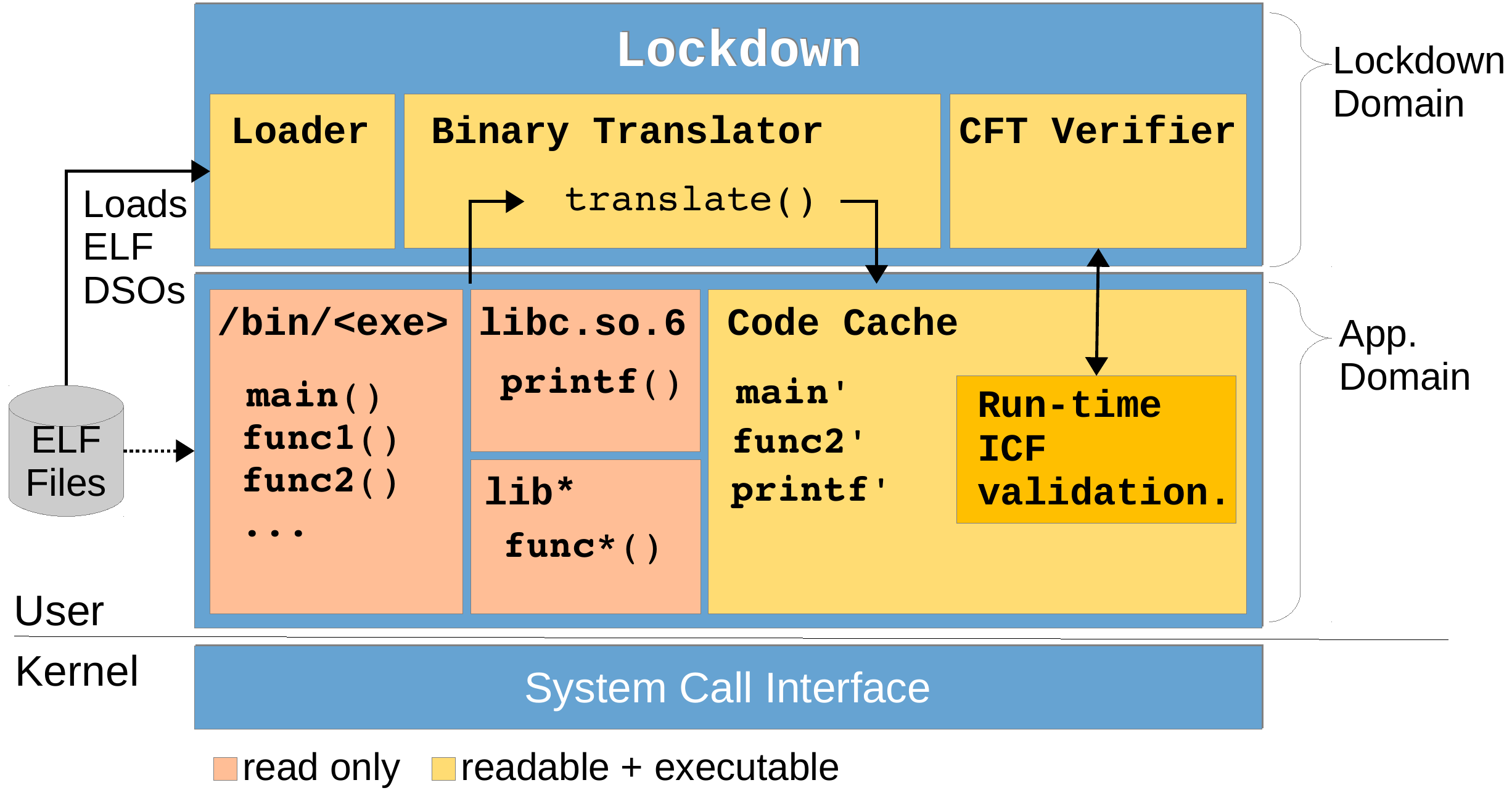}
    \caption{Overview of the Lockdown approach.}
    \label{fig:lockdown}
  \end{center}
\end{figure}

This paper presents Lockdown, a dynamic protection mechanism for legacy,
binary-only code that recovers fine-grained control-flow graph information at
runtime.  Lockdown enforces a strict CFI policy for function calls and indirect
branches and adds a shadow stack to protect the integrity of return instructions
at all times. Compared to other CFI mechanisms, Lockdown (i) relies on a dynamic,
on-the-fly analysis (compared to a-priori static analysis), (ii) enforces a
module-aware dynamic CFI policy that adapts according to the currently loaded
libraries (function calls are restricted to valid functions inside the same
module or correctly imported functions from other libraries), and (iii)
restricts jump instructions to valid instructions inside the same function.
Lockdown is more precise than existing binary-only CFI protections as CFI is
enforced on a per-library granularity. When available, Lockdown uses
non-stripped libraries to infer information about internal symbols.
In addition, Lockdown uses a set of simple
heuristics to detect callback functions that are not exported but passed between
modules, to protect exceptions, and to support tail recursion implementations.
Lockdown relies on a trusted loader~\cite{secuLoader} to extract runtime
information about loaded libraries and possible interactions between those
libraries and uses a sandbox based on dynamic binary translation~\cite{libdetox}
to enforce the integrity of individual control-flow transfers.
\autoref{fig:lockdown} shows an overview of Lockdown.

Our open-source prototype implementation of Lockdown for x86 Linux protects
arbitrary applications from control-flow hijack attacks by implementing our
dynamic CFI policy. An evaluation of the performance of the prototype
implementation based on the SPEC CPU2006 benchmark shows reasonable performance overhead. 
In addition, we introduce the concept of ``dynamic average indirect target reduction''
to allow an empirical evaluation of the resilience of
Lockdown-protected applications against control-flow hijack attacks.

This paper makes the following contributions:
\begin{enumerate}
\item Design of Lockdown, a security mechanism that enforces dynamic precise
control-flow integrity by reconstructing a runtime control-flow graph using
auxiliary data available through the dynamic loader;
\item A security evaluation of the increased protection that Lockdown offers
using Dynamic Average Indirect target Reduction (DAIR), a dynamic metric that
evaluates the possible targets for a given library at any given time;
\item A case study and performance evaluation of Lockdown and our prototype
implementation using a set of common applications. %
\end{enumerate}

The rest of the paper is organised as follows: \autoref{sec:attackmodel}
discusses the attack model followed by related work in \autoref{sec:rw};
\autoref{sec:design} presents the design of Lockdown, \autoref{sec:impl}
discusses the implementation, and \autoref{sec:eval} evaluates the prototype
implementation; and \autoref{sec:conclusion} concludes.

\section{Attack model}\label{sec:attackmodel}

For Lockdown we assume a powerful yet very realistic attack model: the attacker
has read and write capabilities of the program's data regions and read
capabilities for the program's code regions. This attack model reflects common
efforts to circumvent the deployed defenses on current systems.  The attacker
uses memory corruption vulnerabilities present in the application or any of the
loaded libraries to modify the program's data, thereby affecting the
control-flow and execution of the program. 
The attacker leverages either out-of-bounds pointers or dangling pointers to
read/write data in the program's address space, with the result of forcing
behavior undefined in the source programming language.

We assume that the attacker can neither modify the code region of the
program nor inject additional code into the program. This assumption is
fulfilled by current systems that enforce an W$\oplus$X~\cite{execshield}
strategy, where any memory area is either writable or executable 
(and never both at the same time). To achieve code execution 
capabilities the attacker must therefore reuse existing code sequences
available in some code region of the program or its libraries. In code reuse
attacks~\cite{ret2libc, nergal07ret2libc, beyondstacksmashing, bletsch11asiaccs,
beyondstacksmashing, bittau14sp, bosman14sp} the attacker prepares a set of
invocation frames that point to code sequences (so called ``gadgets'') that end
with an indirect control-flow transfer whose target is again controlled by the
attacker.

Using these given (practical) capabilities an attacker will try to (i) overwrite
a code pointer, (ii) prepare a set of invocation frames for a code reuse attack,
and (iii) force the program to dereference and follow the compromised code
pointer.

\section{Background and related work}\label{sec:rw}

A variety of defense mechanisms exist that protect against control-flow hijack
attacks by protecting the integrity of code pointers (see Szekeres et
al.~\cite{szekeres13oakland} for a systematization of attacks and defense
mechanisms). Existing defense mechanisms stop the control-flow hijack attack at
different stages by: (i) retrofitting type safety and/or memory safety onto
existing languages~\cite{necula2005ccured, jim02atc}, (ii) protecting the
integrity of code pointers (i.e., allowing only valid code locations to change
the memory area of a code pointer)~\cite{nagarakatte09pldi, nagarakatte2010ismm,
akritidis08sp}, (iii) randomizing the location of code regions or code blocks
(ASLR or code diversification are examples of this probabilistic
protection)~\cite{aslr, propolice01hiroaki}, or (iv) verifying the correctness
of code pointers when they are used. E.g., Control-Flow Integrity
(CFI)~\cite{ccs05erlingsson} is a defense mechanism that stops the attack in the
fourth stage, by preventing the use of a corrupted code pointer.  Unfortunately,
the implementation of CFI is challenging and consequently, CFI has not yet seen
widespread use. 

Software-based fault isolation (SFI) protects from faults in the application and
enforces a given security policy on an unaware program. SFI can be implemented
using a dynamic binary translation system. 

Lockdown combines a dynamic binary translation system with a trusted loader to
enforce a dynamic, per-module CFI policy on unaware, binary-only programs using
the information about available modules and functions from the trusted loader.

\subsection{Control-Flow Integrity}

Control-Flow Integrity (CFI)~\cite{ccs05erlingsson} and its extension
XFI~\cite{osdi06erlingsson} restrict the control-flow of an application at
runtime to a statically determined control-flow graph. Each indirect
control-flow transfer (an indirect call, indirect jump, or function return) is
allowed to transfer control at runtime only to the set of statically determined
targets of this code location. 

CFI relies on code integrity (i.e., an attacker cannot change the executed code
of the application). As explained in \autoref{sec:attackmodel}, the only way an
attacker can achieve code execution is by controlling code pointers. CFI checks
the integrity of code pointers at the location where they are used in the code.
Using memory corruption vulnerabilities an attacker may change the values of
code pointers (or any other data). The attack is detected (and stopped) when the
program tries to follow a compromised code pointer.

The effectiveness of CFI relies on two components: (i) the (static) precision of
the control-flow graph that determines the upper bound of precision and (ii) the
(dynamic) precision of the runtime checks. 

First, CFI can only be as precise as the control-flow graph that is enforced. If
the control-flow graph is too permissive, it may allow an illegal control
transfer. All existing CFI approaches rely on two phases: an explicit static
analysis phase and an enforcement phase that executes additional checks. Most
compiler-based implementations of CFI~\cite{ccs05erlingsson, osdi06erlingsson,
bletsch11acsac, wang10oakland, philippaerts11dimva, akritidis08sp, zeng13usenix,
ben13ccs} rely on a points-to analysis for code pointers at locations in the
code that execute indirect control-flow transfers.  A severe limitation of these
approaches is that all the protected code must be present during compilation as
they do not support modularity or shared libraries.  Implementations based on
static binary analysis~\cite{zhang13oakland, davi12ndss, xi12dsn,
zhang13asiaccs} either rely on relocation information (e.g., in the Windows PE
executable format) or reconstruct that information using static
analysis~\cite{zhang13security}. Modular CFI~\cite{niu14pldi} is a recent
compiler-based CFI tool that stores type information and dynamically merges
points-to sets when new libraries are loaded (but does not support library
unloading).

Second, the initial upper bound for precision is possibly limited through the
implementation of the control-flow checks (Goektas et al.~\cite{goektas14sp}
list common limitations in CFI implementations).
Practical implementations often maintain three global sets of possible targets
instead of one set per control-flow transfer: one target set each for indirect
jumps, indirect calls, and function returns. The control-flow checks limit the
transfers to addresses in this set. This policy is an improvement compared to
unchecked control-flow transfers but overly permissive as an attacker can hijack
control-flow to any entry in the set.

Lockdown is a dynamic approach that enforces a stricter, dynamically constructed
control-flow graph on top of a dynamic sandbox for binaries. The sandbox ensures
code integrity, adds a safe shadow stack that protects against return-oriented
programming attacks~\cite{ret2libc}, and enforces dynamic control-flow checks.
The secure loader protects the GOT and GOT.PLT data structures from malicious
modifications.

\subsection{Dynamic binary translation}

Software-based Fault Isolation (SFI) protects the integrity of the system
and/or data by executing additional guards that are not part of the original
code. Dynamic Binary Translation (DBT) allows the implementation of SFI guards on
applications without prior compiler involvement by adding a thin virtualization
layer between the original code and the code that is actually executed. The DBT
system dynamically enforces security policies by collecting runtime information
and restricting capabilities of the executed code.

Several DBT systems exist with different performance characteristics.
Valgrind~\cite{nethercote08valgrind} and PIN~\cite{luk05pldi} offer a high-level
runtime interface resulting in higher performance costs while
DynamoRIO~\cite{bruening03dynamorio} and libdetox~\cite{libdetox} support a more
direct translation mechanism with low overhead translating application code on
the granularity of basic blocks. %
We build on libdetox
which has already been used to implement several security policies.

A security policy can only be enforced if the translation system itself is
secure. Libdetox splits the user-space address space into two domains: the
application domain and the trusted binary translator domain.  This design
protects the binary translation system against an attacker that can modify the
address space of the running application as the attacker cannot reach the
trusted DBT domain.  Libdetox uses a separate translator stack and separate
memory regions from the running application.  Libdetox enforces the following
properties: (i) no untranslated code is ever executed; (ii) translated code is
executed in the application domain; (iii) no pointer to the trusted domain is
ever stored in attacker-accessible memory. The application triggers a trap into
the trusted domain when (i) it executes a system call, (ii) executes
untranslated code, or (iii) a heavy-weight security check is triggered. Libdetox
can be extended by the addition of a trusted loader to the trusted computing
domain~\cite{secuLoader} thereby protecting the SFI system from attacks against
the loader when loading or unloading shared libraries.

The combination of trusted loader and dynamic binary translation system
implements the following security guarantees: a \emph{shadow stack} protects the
integrity of return instruction pointers on the stack at all times; the
\emph{trusted loader} protects the data structures that are used to execute
functions in other loaded libraries at runtime; and the \emph{integrity} of the
security mechanism is guaranteed by the binary translation system. 
The shadow stack is implemented by translating call and return
instructions~\cite{libdetox}. Translated call instructions push the return
instruction pointer on both the application stack and the shadow stack in the
trusted domain. Translated return instructions check the equivalence between the
return instruction pointer on the application stack and the shadow stack; if the
pointers are equivalent then control is transferred to the translated code block
identified by the code pointer on the shadow stack.

The existing version of libdetox supports and allows the application to execute
the full x86 instruction set (including any SSE extensions). The trusted
computing base of the architecture (trusted loader and binary translation
system) is small, with less than 20,600 lines of code.

\subsection{Dynamic loading in a nutshell}

Modern Unix operating systems and the Linux kernel use the Executable and
Linkable Format~\cite{sysVABI,dreppersharedlibs} (ELF) to specify the on-disk
layout of applications, libraries, and compiled objects. A file that uses the
ELF format to define its internal layout is called Dynamic Shared Object
(DSO). The ELF format defines two views for each DSO. The first view is the
program header that contains information about segments; the program header
controls how the segments must be mapped from disk into the process image. The
second view is the section header table; this table contains the more detailed
section definitions.

The dynamic loader and the linker both use the section header table. Depending
on the linker flags the section header table of a linked executable or shared
library contains information about all functions and symbols, or it contains
only information about exported symbols.

The \texttt{dynsym} section contains the location information of all exported
symbols and is available in all dynamically linked libraries and
applications. This section is used by the dynamic loader to resolve references.
The \texttt{symtab} section on the other hand is not needed by the dynamic
loader and can be stripped from the final library or application (but is
available for most ELF files on current systems). This section contains detailed
information at the highest possible level of granularity and includes details of
all available symbols, even static and weak variables or functions, hidden
symbols, and individual object files. The granularity of the \texttt{dynsym} and
\texttt{symtab} information is no longer on the section level but on the level
of individual compiled objects (usually individual source files) and functions.
If available this information is used, e.g., in the debugger.

Lockdown uses the \texttt{dynsym} information to establish an information
baseline. This base is extended with the more detailed \texttt{symtab}
information to form the complete picture using as many details as possible. If
the \texttt{symtab} information is not available, e.g., because the binary is
stripped, then Lockdown falls back to the information baseline.

\section{Lockdown design}\label{sec:design}

\begin{figure*}
  \begin{center}
    \includegraphics[width=120mm]{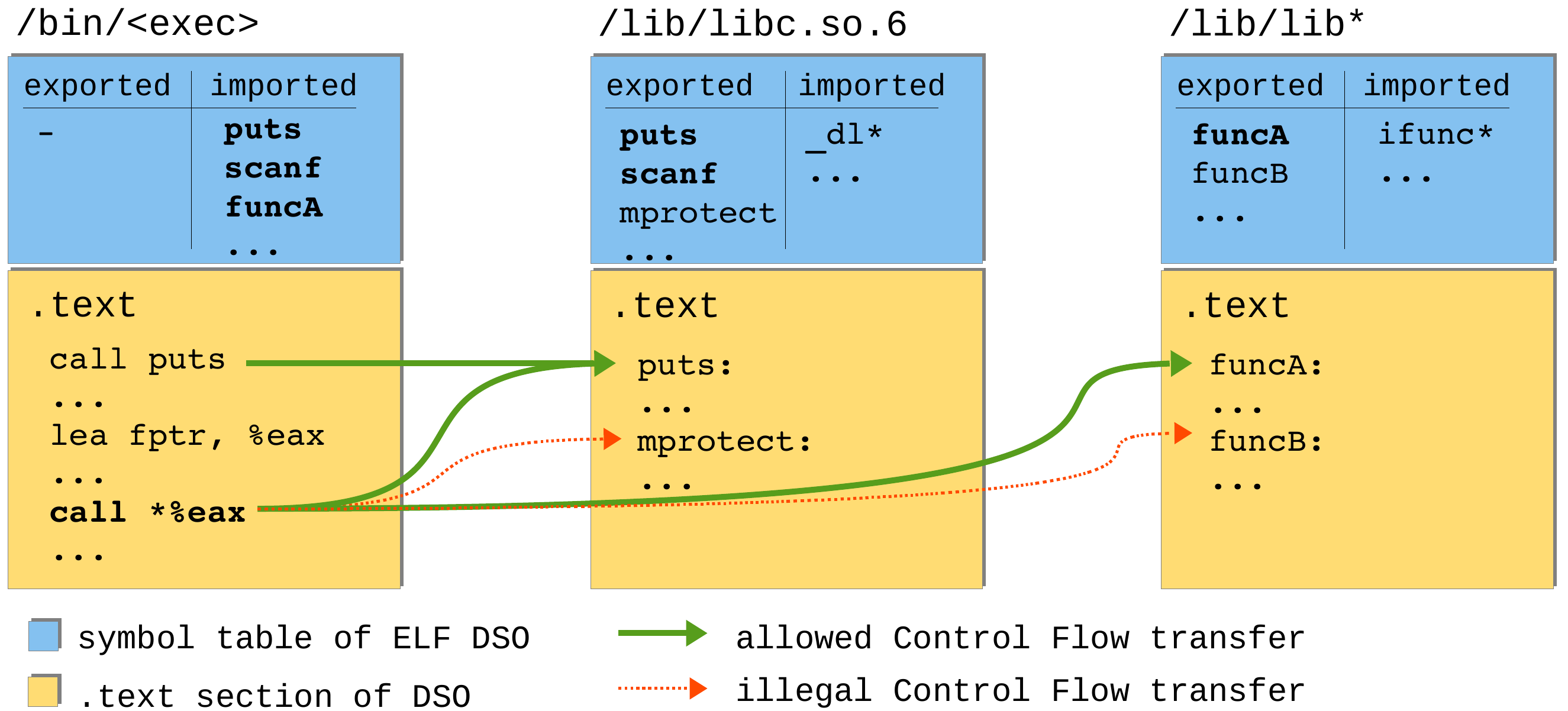}
    \caption{Call restrictions for an executable and two libraries.
    Executables and libraries are only allowed to call imported 
    function symbols. Local function calls may only transfer to local function
		symbols.}
    \label{fig:call-restrictions}
  \end{center}
\end{figure*}

Lockdown enforces a strict, practicable CFI policy at the level of shared
libraries by restricting control-flow transfers between loaded shared libraries.
The high-level policy restricts (i) inter-module calls to functions that are
exported from one library and imported in the other library, (ii) intra-module
calls to valid functions, (iii) jump instructions to valid instructions in the
same function and valid call targets for tail calls, and (iv) return
instructions to the precise return address (with a special handler for
exceptions). \autoref{fig:call-restrictions} shows an example of the call
restrictions for three loaded libraries. The call restrictions are adapted
dynamically whenever libraries are loaded or unloaded.
The integrity of return instructions is enforced at all times using a shadow
stack. Indirect call instructions and indirect jump instructions execute a
runtime check that validates the current target according to the currently
loaded libraries.

Lockdown relies on a trusted loader that prepares information about all visible
symbols relative to each code location. The loader uses the available symbol
tables and information about imported and exported symbols to provide this
information. If a shared library is stripped then only information about
imported and exported symbols is available, resulting in a coarser-grained
protection. 

\newpage
\subsection{Rules for control transfers}\label{sec:rules}

Three different forms of control transfers exist, \texttt{call}, \texttt{jump},
and \texttt{return} instructions. Call instructions are used to transfer control
to a different function. Jump instructions (conditional and unconditional), on
the other hand, are used to transfer control inside a function. Return
instructions are used to transfer control back to the calling function. The
rules for different control transfers are as follows:

\begin{enumerate}
  \item Call instructions must always target valid functions. The set of valid
    functions is always specific for every protected call instruction. Only
    functions that are defined or imported in the current module are allowed.
    Static functions or local functions that are called in the same module are
    only verifiable if the binary is not stripped.
  \item \label{jmprule} The target of a jump instruction must stay within the same
    function or must go to a valid function target (due to tail call
		implementation). %
    If the binary is stripped then jumps can only be verified to the granularity
    of individual sections and exported symbols.
  \item Return instructions must always transfer control back to the caller. The
		DBT system keeps a shadow stack data structure in the trusted domain to
    verify return addresses on the stack before they are dereferenced. The
    shadow stack verifies the equivalence of the return address on the
    application stack and removes it from the application stack. It then removes
    the return address from the shadow stack and transfers control to the
    translated counterpart of that code pointer (protecting from time of check
		to time of use attacks).%
  \item All control transfers must target valid instructions that are
    reachable from the beginning of a symbol. This rule prohibits control
    transfers into instructions.%
\end{enumerate}

These rules hold for correctly compiled programs under the assumption that an
attacker cannot modify code pointers in memory. Under our attack model in
\autoref{sec:attackmodel} the attacker can freely modify any code pointer in
writable memory. Lockdown inserts additional guards that protect control-flow
transfers and detects code pointers that were modified by an attacker when that
code pointer is used. If an individual control transfer breaks one of the rules
above then the application is terminated with a security exception.

\subsection{Control transfer categories}

Most instruction sets (including x86) allow two types of control transfers,
namely static control transfers that redirect control-flow to a target that is
encoded in the instruction itself and indirect control-flow transfers that
transfer control indirectly through a memory address (the code pointer) that
contains the target. An indirect control-flow transfer encodes the memory
address or register that contains the target in its instruction.

All static control-flow transfers are verified during the translation of the
instruction. %
Because all information is static this translation-time check is
safe and holds as long as only new code is added to the cache of translated
code. If library code is removed (e.g., a shared library is unloaded), then the
bindings to the removed code are revoked, and the code is revalidated (at the
cost of a new translation).

Indirect control-flow transfers on the other hand are highly dynamic. The
instruction points to a memory address (which is under the control of the
attacker) that contains the real target of the control-flow transfer. The
contents of the memory address can change every time the instruction is executed
(e.g., think of a function pointer that is changed every time before it is
executed). A static check during the translation of the instruction does not
suffice to verify these indirect transfers. The binary translator must emit
an additional guard that executes a dynamic check every time the dynamic
control-flow is dispatched. %

Return instructions mark the end of a function and return control from the
called function back to the caller. This control transfer is also a form of
indirect control transfer. The target of the return instruction is usually
(i.e., if no exception is thrown) determined by the last call instruction in the
instruction stream. The call instruction pushes the current instruction pointer
onto the stack and transfers control to the called function. The return
instruction on the other hand pops the topmost location from the stack and
transfers control back to the caller.  Return instructions are translated into a
special sequence of instructions. The added guard verifies that the current
return address on the application stack is the same as the address that was
pushed by the last call instruction.

\subsection{Control transfer lookup table}\label{ctlt}

Both dynamic guards and static checks use a \textit{control transfer lookup
table} to verify that the actual target is allowed for the current control-flow
transfer. This dynamic data structure maps target locations to a set of possible
source locations. The source locations are either based at the granularity of a
dynamically shared object (e.g., a shared library or the application itself), an
(dynamic) object, or section inside the shared library, a function in a section,
or a specific memory location.

The trusted loader constructs the control transfer lookup table on the fly. The
table is updated whenever the dynamic loader resolves a new dynamically shared
object. This table keeps information about all loaded symbols and functions and
possible regions that can access this function. If the compiled ELF DSO contains
the full symbol table then it is possible to determine the locations of all objects
and all functions. If the binary is stripped and there is only dynamic loader
information available then only the locations of exported symbols are known, and
there is no information about the location of individual static functions and
objects. For most executables and libraries the full symbol table is available
(or can be installed using a separate package). If (e.g., legacy) objects
are stripped then the precision of Lockdown is limited to the set of exported functions.

\section{Prototype implementation}\label{sec:impl}

The prototype implementation builds on libdetox and seculoader~\cite{secuLoader,
libdetox}. We implement the control-flow checks in the DBT during the
translation of individual basic blocks. All static control-flow transfers are
verified during the translation and indirect (dynamic) control-flow transfers
are instrumented to execute an inlined reference monitor that executes a dynamic
guard depending on the type of control-flow transfer.

Our current prototype does not support self-modifying code. Given our attack
model we cannot decide if new code was produced by the attacker or a
just-in-time compiler that is a part of the application. We therefore terminate
the application if we detect code generation or self-modifying code.  (It would
be possible to allow self-modifying code if the JIT compiler is part of the
trusted computing base)

For all DBT systems one of the biggest performance overheads is the translation
of indirect control-flow transfers. These indirect control-flow transfers cannot
be translated ahead of time and always incur a runtime lookup to consult the
mapping between the original code and the translated code blocks, and only then
control-flow is transferred to the translated code block. These additionally
executed instructions lead to performance overhead and we use a lookup cache to
reduce this overhead.  Caching is an optimization that locally stores the last
successful pair of original code pointer and translated code pointer. If the
control-flow check was successful and the current target is still the same then
there is no need to execute the control-flow check again.

The prototype implementation is released as open-source and was implemented
using 22,000 lines of C code mixed with a small amount of inline assembler
instructions.

\subsection{Runtime optimizations}

Achieving low overhead when running binary-only applications is a challenging
problem: to support dynamic CFI policies Lockdown needs to run the binary
analysis alongside the executing application. For our prototype implementation
we have implemented a set of optimizations to achieve low overhead without loss
of precision or security.
The libdetox DBT engine already implements a set of optimizations like local
inline caches for indirect control-flow transfers. We extend the indirect
control-flow transfer lookups by a control-flow transfer check. The validation
is split into a fast path and a slow path. The fast path uses a lookup table for
already verified \texttt{\{sourcedso,destination\}} pairs. The slow path
recovers if this fast check fails (e.g., if there is a hash miss in the lookup
table).
Lockdown handles C++ exceptions by unwinding and resynchronizing the shadow
stack with the application stack. The stack handling routine removes frames from
the shadow stack until the frames match again. Shadow stack frames can only be
removed by exceptions (they are never added), resulting in sound behavior.

ELF implements calls to other libraries as a call to the PLT section of the
current module and an indirect jump to the real function. The DBT replaces such
calls with a direct call to the loaded function, removing the indirect jump and
needed CFI guard. In addition, the trusted loader protects the PLT and PLT.GOT
data structures from adversarial access.

An optimization that we have left for future work replaces the global cache of
translated code with a per-module cache. This allows to statically encode and
protect inter-module transfers, removing the need to execute an expensive CFI
guard if the function pointer points into the same module.

\subsection{Control-flow particularities}

Control-flow transfers in off-the-shelf binaries do not always adhere to the
rules listed in \autoref{sec:rules} and Lockdown catches this behavior and
recovers using a set of handlers. These special cases are specific to low level
libraries like the libc run-time support functions, e.g., inter-module calls to
symbols that were not imported, intra-module cross function jumptables,
inter-module callback functions or even inter-module calls targeting PLT entries
which would bypass our PLT inlining if not handled correctly. Lockdown also
allows indirect jumps as tail calls to the beginning of other functions in the
set of currently allowed call targets. Although a variety of these special cases
exist they can all be handled by a small set of handlers without compromising
the CFI security properties.

High-level application code (i.e., all code that is not the libc or other
low-level functionality like the loader) adheres to the rules listed in
\autoref{sec:rules} and therefore does not require special handling. The only
exception that needs global special handling are callback functions which are
discussed in the next section.

\subsection{Implementation heuristics}

Binaries have little information about the types that are used at runtime and it
is not always possible to recover information precisely. To support callback
functions (i.e., a function in a library returns a function pointer which is
later called from a different library; if this function is not exported/imported
then the CFI guard would fail) Lockdown implements a dynamic scanning technique
that is similar in design to the static analysis of Zhang and Sekar proposed
in~\cite{zhang13security}. Their system statically checks for instruction
sequences that are used to calculate function pointers. 

Lockdown uses the following patterns to detect callback pointers to callback
functions on the fly (i) \texttt{push imm32} where a function pointer is pushed
onto the stack, (ii) \texttt{movl imm32, rel(\%esp)} where rel references a
local variable on the stack, and (iii) \texttt{leal imm32(\%ebx), \%e?x} where a
function pointer is moved from memory into a general purpose register relative
to .got.plt, or (iv) relocations that are used to define pointers for many
callbacks (e.g., \texttt{R\_386\_RELATIVE}). An advantage of Lockdown's dynamic
analysis compared to static analysis is that Lockdown uses the actual (precise)
values at runtime and checks whether the pointer actually references a valid
instruction. Lockdown is therefore potentially more precise than a static
analysis as static analysis is limited if there are cross-module control-flow
transfers that are rely on complicated (dynamic) control-flow patterns and
computation that spans across several basic blocks that are connected through
indirect control-flow transfers (and maybe even across shared libraries). In
addition, Lockdown scans the \texttt{.data} region for relocations pointing into
the code segment to detect static code pointers.

The trusted loader protects all runtime loader data structures from adversarial
access and implements both the \texttt{dlopen} class of functions and direct
data structure access that libc uses.
The trusted loader component allows an interesting way to deploy our dynamic CFI
system: in the ELF file any executable may specify the default loader that is
used to execute it. Lockdown is deployed on a per-application basis by replacing
this default loader with the Lockdown executable, thereby replacing the standard
loader with Lockdown.

\section{Evaluation}\label{sec:eval}

We evaluate Lockdown in the following areas: (i) performance using the SPEC
CPU2006 benchmarks, (ii) real-world performance using Apache
2.2, %
(iii) a
theoretical discussion of the security guarantees according to the implemented
security policy, and (iv) an evaluation of the remaining attack surface. 

We run the experiments in the following sections on an Intel Core i7 CPU
920@2.67GHz with 12GiB memory on Ubuntu 12.04.4. Lockdown and the SPEC CPU2006
benchmarks %
are compiled with gcc version 4.6.3. Apache 2.2.22 is installed from the
default package. The full set of security features of Ubuntu 12.04.4 is enabled
(ASLR, DEP, stack canaries, and safe exception frames).

\subsection{Performance}

We use the SPEC CPU2006 benchmarks to measure raw CPU performance and to
evaluate the performance impact of Lockdown with full indirect control-flow
protection compared to native execution and binary translation (with shadow
stack) only execution.

\autoref{tbl:spec} shows the performance results for SPEC CPU2006 for
native, binary translation only and Lockdown runs. Due to issues with the
trusted loader we were unable to run omnetpp and dealII. The binary translation column
already includes the shadow stack from libdetox~\cite{libdetox} and overhead
from the trusted loader. The additional
average overhead introduced by Lockdown for CFI enforcement (compared to binary
translation) is 17.37\%, and the average overhead including binary translation is
32.49\%. Only five benchmarks have a total overhead of more than 45\% if run in
Lockdown. The majority of benchmarks face a performance overhead of around 20\%
or below. Overall the overhead for dynamic CFI enforcement is reasonable. In
future work we are looking into reducing the overhead by more aggressive caching
and inlining.

\begin{table*}[t!]
\begin{center}
  \begin{tabular}{  l | r || r | r || r | r }
  benchmark & native & BT only & overhead & Lockdown & overhead \\ \hline
  400.perlbench & 408 & 848 & 107.84\% & 1522 & 273.04\%\\
  401.bzip2 & 693 & 741 & 6.93\% & 740 & 6.78\%\\ 
  403.gcc & 360 & 511 & 41.94\% & 666 & 85\%\\
  429.mcf & 300 & 309 & 3\% & 307 & 2.33\%\\
  445.gobmk & 535 & 736 & 37.57\% & 767 & 43.36\%\\
  456.hmmer & 617 & 636 & 3.08\% & 633 & 2.59\%\\ 
  458.sjeng & 643 & 1014 & 57.7\% & 1454 & 126.13\%\\ 
  462.libquantum & 562 & 579 & 3.02\% & 580 & 3.2\%\\
  464.h264ref & 828 & 1055 & 27.42\% & 1738 & 109.9\%\\
  473.astar & 548 & 601 & 9.67\% & 739 & 34.85\%\\
  483.xalancbmk & 289 & 570 & 97.23\% & 962 & 232.87\%\\
  410.bwaves & 589 & 587 & -0.34\% & 599 & 1.7\%\\
  416.gamess & 1104 & 1212 & 9.78\% & 1333 & 20.74\%\\
  433.milc & 518 & 545 & 5.21\% & 558 & 7.72\%\\
  434.zeusmp & 627 & 626 & -0.16\% & 626 & -0.16\%\\
  435.gromacs & 1002 & 1025 & 2.3\% & 1026 & 2.4\%\\ 
  436.cactusADM & 1207 & 1244 & 3.07\% & 1249 & 3.48\%\\
  437.leslie3d & 586 & 588 & 0.34\% & 591 & 0.85\%\\
  444.namd & 578 & 588 & 1.73\% & 586 & 1.38\%\\ 
  450.soplex & 303 & 339 & 11.88\% & 367 & 21.12\%\\ 
  453.povray & 273 & 411 & 50.55\% & 621 & 127.47\%\\
  454.calculix & 1124 & 1152 & 2.49\% & 1168 & 3.91\%\\ 
  459.GemsFDTD & 551 & 576 & 4.54\% & 578 & 4.9\%\\ 
  465.tonto & 712 & 858 & 20.51\% & 869 & 22.05\%\\ 
  470.lbm & 382 & 388 & 1.57\% & 383 & 0.26\%\\ 
  481.wrf & 970 & 1086 & 11.96\% & 1084 & 11.75\%\\ 
  482.sphinx3 & 558 & 592 & 6.09\% & 601 & 7.71\%\\ 
  \hline
  average &  &  & 15.12\% &  & 32.49\%\\
  \end{tabular}
	\caption{SPEC CPU2006 results for native, binary translation only, and
  Lockdown.} 
	\label{tbl:spec} 
\end{center}
\end{table*}

\subsection{Apache case study}

In this section we evaluate the performance of a full Apache 2.2 setup running
under the dynamic CFI protection of Lockdown (including ROP protection using the
shadow stack). Apache is set up in the default configuration and we use the ab
Apache benchmark to measure performance.

To test the performance of the web server we use an html file (56 KB)
and a jpg image (1054 KB) that are served by Apache. The file sizes were chosen
according to the average html and image sizes reported by~\cite{httparchive}.
We used ab to send 5,000,000 requests and measured the overall time required
to respond to these requests.

The average overhead of Apache 2.2 running inside Lockdown is 33\% with an
overhead of 57\% for the small html file and 10\% for the larger jpg file. The
additional context switches between translator domain and application domain for
file and network I/O operations become more dominant for smaller files. Apache
sends files using as few I/O operations as possible and with small files there
is not enough computation that is executed to recover from the performance hit
of the context switch.

\subsection{Security evaluation}

Evaluating the effectiveness of a CFI implementation in terms of security is not
trivial. Running a vulnerable program with a CFI implementation and preventing a
specific exploitation attempt does not mean that the vulnerability is not
exploitable under other circumstances or by hijacking control-flow along other
paths that actually might be allowed within the control-flow graph of the CFI
implementation.

We therefore make the following observations: (i) in our attack model a
successful attacker needs to hijack the control-flow to already executable code
within the process, (ii) the probability of success for an attacker depends on
the ability to find a sequence of reusable code (gadgets) that executed in the
right order accomplishes the intended malicious behaviour (e.g. running a shell).

The effectiveness of a CFI implementation therefore depends on how effectively
an attacker is restrained in the ability to find and reuse already available
code. This directly translates to the quantity and quality of the still
reachable indirect-control flow (ICF) targets of the enforced CFG.

Zhang and Sekar~\cite{zhang13security} propose a metric for measuring CFI
strength called Average Indirect target Reduction (AIR). Based on AIR we define
the Dynamic Average Indirect target Reduction (DAIR):

{\bf Definition: Dynamic Average Indirect target Reduction}
{\it Let t be a specific point in time during a program's execution and
$i_{1}$,...,$i_{n}$ be all the ICF transfers in a program executed before and at time t.
A CFI technique limits possible targets of ICF transfer $i_{j}$ to the set
$T_{j}$. We define DAIR(t) as:}
\begin{displaymath}
{\bf DAIR(t)} = \frac{1}{n}\sum_{j=1}^{n} (1 - \frac{|T_j|}{S})
\end{displaymath}
{\it where S is the number of possible ICF targets in an unprotected program.}

This definition of DAIR is almost identical to AIR defined by Zhang and Sekar
except that it is dynamic and takes into account only ICFs that where
executed during and up to a certain point in time $t$ of a program's execution.
Therefore the DAIR metric is not static and  varies during program execution.
DAIR allows us to quantify the effectiveness of Lockdown's CFI implementation
despite its dynamic nature.

In addition to the quantity of ICF targets that are still valid, it is also
necessary to consider the ``quality'' of the remaining ICF targets
(with regard to the target's ability to attack successfully). Unfortunately,
there is not one unique type of instruction sequence that might be used by an attacker
-- depending on what an attacker wants to achieve, the criteria to assess a target's
quality vary depending on the attacker's gadget's needs.
It is therefore difficult to measure CFI effectiveness based on a quality metric for 
ICF targets and their usefulness to an attacker. We therefore consider DAIR
as the sole metric for CFI strength.

\autoref{tbl:dair} shows the DAIR values for Apache, the SPEC CPU2006 benchmarks,
and a set of common applications. Comparing
the final DAIR value at program termination (i.e., the most open value that
includes all values discovered by Lockdown) with the AIR values presented by
Zhang and Sekar we show an improvement of several percent. In addition, the
return instruction pointers on the stack can only be redirected to one (of the
few) pointers further down on the stack, stopping ROP attacks.

\begin{table*}[t!]
\begin{center}
    \begin{tabular}{  l | c || c | c | c }
     DAIR & {\bf total } & ind. call & ind. jump & return \\ \hline
     ls & 99.90\% & 99.71\% & 99.05\% & 99.99\% \\ 
     hostname & 99.95\% & 99.81\% & 99.91\% & 99.99\% \\ 
     netstat & 99.67\% & 98.13\% & 99.28\% & 99.99\% \\ 
     nano & 99.96\% & 99.78\% & 99.70\% & 99.99\% \\ 
     vim & 99.21\% & 94.39\% & 97.29\% & 99.99\% \\ 
     nmap & 99.70\% & 99.08\% & 93.04\% & 99.99\% \\
     xmessage & 99.67\% & 97.41\% & 99.60\% & 99.99\% \\ 
     xcalc & 99.72\% & 97.69\% & 99.71\% & 99.99\% \\
     xterm & 99.77\% & 98.48\% & 98.08\% & 99.99\% \\
     apache & 99.58\% & 97.65\% & 96.90\% & 99.99\% \\
    \hline
     average & {\bf 99.71\% } & 98.21\% & 98.26\% & 99.99\% \\
    \hline
     SPEC CPU 2006 & {\bf 99.93\% } & 99.81\% & 99.33\% & 99.99\% \\
    \end{tabular}
    \caption{DAIR of a small set of UNIX applications, Apache and the SPEC CPU 2006 benchmarks, at time t = program
    termination time. Where available, debug symbol files of libraries were used.}
	\label{tbl:dair}
\end{center}
\end{table*}

\autoref{tbl:dair} shows the DAIR values for a set of UNIX applications at
t = "program termination". The DAIR values reflect how much, on
average, the executed ICF instructions' set of valid targets is
reduced. We see that Lockdown is very effective for ret instructions as the
shadow stack allows us to restrict potential ret targets to their exact value.
As expected Lockdown is slightly more effective for indirect calls than for indirect jmps
as the CFG is more granular at inter-module boundaries.
In general, Lockdown achieves very high DAIR values as the set of valid targets
is specific per loaded code module. In comparison, static CFI implementations have AIR
values of 99.13\% (CFI reloc) or 98.86\% (CFI bin) as reported by Zhang and Sekar~\cite{zhang13security}.
Directly comparing raw AIR with DAIR values is not feasible as the DAIR view
changes over time. We compare the final (most permissive) DAIR numbers at
program termination with the AIR number to give an idea of how Lockdown's CFI
strength compares to other static CFI approaches.

\autoref{tbl:dairstripped} shows the DAIR values when only stripped libraries are used.
Lockdown still achieves reasonable CFI strength despite the lack of extended
symbol information. Fortunately, almost all Linux distributions provide a way to
install symbol file packages. For the values in \autoref{tbl:dair} we used the
symbol file packages of Ubuntu 12.04.4 for common libraries.

\begin{table*}[t!]
\begin{center}
    \begin{tabular}{  l | c || c | c | c }
     DAIR & {\bf total } & ind. call & ind. jump & return \\ \hline
     UNIX average & {\bf 95.21\% } & 72.75\% & 83.52\% & 99.99\% \\
     SPEC CPU 2006 & {\bf 95.39\% } & 74.07\% & 86.42\% & 99.99\% \\
    \end{tabular}
    \caption{DAIR average of a small set of UNIX applications and the SPEC CPU 2006 benchmarks, at time t = program
    termination time. All libraries are {\bf stripped}. %
		}
	\label{tbl:dairstripped}
\end{center}
\end{table*}

As mentioned before the quantity of the ICF target reduction covers only one aspect.
Lockdown's advantage to other CFI approaches is that valid targets are specific
to individual ICF instructions. This is especially relevant considering powerful functions
like {\bf mprotect()} or {\bf system()}. These libc functions are interesting for
attackers because of their high misuse potential. Lockdown will deny calls to these
functions from DSOs that do not explicitly import these symbols. Therefore even if one
module in a process is allowed to call them other modules will still be restrained in
using them.

\subsection{Security guarantees}
Lockdown enforces strict (modular) security guarantees for executed code: (i)
the DBT always maintains control of the control-flow, (ii) only valid, intended
instructions are executed, (iii) function returns cannot be redirected,
mitigating ROP attacks, (iv) jump instructions can target only valid
instructions in the same function or symbols in the same module, (v) call
instructions can target only valid functions in the same module or imported
functions, (vi) all signals are caught by the DBT system, protecting from signal
oriented programming, (vii) all system calls go through a system call policy
check.  Due to the modular realisation, individual guarantees build on each
other: the binary translator ensures the SFI properties that only valid
instructions can be targeted, the shadow stack protects return instructions at
all times, the trusted loader provides information about allowed targets for
call and jmp instructions, and the dynamic control-flow transfer checks enforce
dynamic CFI.

The remaining attack surface is vastly reduced: the GOT.PLT sections are never
writeable, return instruction pointers on the stack are always protected using
the shadow stack, and each remaining indirect control-flow transfer has a
restricted, individual (on a per-module level) set of allowed targets.

\section{Conclusion}\label{sec:conclusion}

This paper presents Lockdown, a strong dynamic control-flow integrity policy for
binaries. Using the symbol tables available in shared libraries and
executables we build a control-flow graph on the granularity of shared objects. A
dynamic binary translation based system enforces the integrity of control-flow
transfers at all times according to this model. In addition, Lockdown uses a
shadow stack that protects from all ROP attacks.

Our prototype implementation shows reasonable performance overhead of 32.49\% on
average for SPEC CPU2006. In addition, we have reasoned about CFI effectiveness
and the strength of Lockdown's dynamic CFI approach which is more precise than
other CFI solutions that rely on static binary rewriting.

Lockdown enforces strong security guarantees for current systems
in a practical environment that allows dynamic code loading (of shared
libraries), supports threads, and results in low overhead.

\newpage
\bibliographystyle{abbrv}
\bibliography{bibliography}  %
\end{document}